\documentclass[reprint,prc,twocolumn,showpacs,amsmath,amssymb,aps]{revtex4-1}

\usepackage{bm} 
\usepackage{hyperref} 
\usepackage{natbib} 
\usepackage{subfig} 
\usepackage{graphicx} 

\newcommand{\sdata}{$N_{\odot,r}$} 
\newcommand{\nggn}{$(n,\gamma)\rightleftarrows(\gamma,n)$}  
\newcommand{\taungtaugn}{$\tau_{n\gamma}/\tau_{\gamma n}$} 
\newcommand{\taubtaung}{$\tau_{\beta}/\tau_{n\gamma}$}     
\newcommand{\reptaungeqb}{$\tau^{REP}_{n\gamma}=\tau^{REP}_{\beta}$}     
\newcommand{\reptaungltb}{$\tau^{REP}_{n\gamma}\lesssim\tau^{REP}_{\beta}$}     
\newcommand{\reptaungafewb}{$\tau^{REP}_{n\gamma}\approx\text{ a few }\tau^{REP}_{\beta}$}     

\begin{document}

\preprint{APS/123-QED}

\title{Formation Of The Rare Earth Peak: Gaining Insight Into Late-Time r-Process Dynamics}

\author{Matthew R. Mumpower}
\email{mrmumpow@ncsu.edu}
\affiliation{Department of Physics, North Carolina State University, Raleigh, North Carolina 27695-8202,USA}

\author{G. C. McLaughlin}
\email{gail_mclaughlin@ncsu.edu}
\affiliation{Department of Physics, North Carolina State University, Raleigh, North Carolina 27695-8202,USA}

\author{Rebecca Surman}
\email{surmanr@union.edu}
\affiliation{Department of Physics and Astronomy, Union College, Schenectady, New York 12308,USA}

\date{\today}

\begin{abstract}
We study the formation and final structure of the rare earth peak ($A\sim160$) of the $r$-process nucleosynthesis. The rare earth peak forms at late times in the $r$-process after neutron exhaustion (neutron-to-seed ratio unity or $R=1$) as matter decays back to stability. Since rare earth peak formation does not occur during \nggn \ equilibrium it is sensitive to the strong interplay between late time thermodynamic evolution and nuclear physics input. Depending on the conditions the peak forms either because of the pattern of the neutron capture rates or because of the pattern of the separation energies. We analyze three mass models under different thermodynamic conditions. We find that the subtleties of each mass model, including separation energies and neutron capture rates, influence not only the final shape of the peak but also when it forms. We identify the range of nuclei which are influential in rare earth peak formation.
\end{abstract}


\maketitle

\newpage

\section{Introduction}\label{intro}
Approximately half of the elements beyond $A>100$ are made in the \textquoteleft rapid\textquoteright\ neutron capture process, or $r$-process, in which successive neutron captures occur on timescales faster than $\beta$-decays. At the present time, there is significant uncertainty with the astrophysical environment responsible for this synthesis event \cite{Arnould200797,Qian2007237}. The leading candidate site \cite{argast2004} is believed to be core-collapse supernovae e.g \cite{1992ApJ...399..656M,1996ApJ...471..331Q,2000PASJ...52..601S,2000ApJ...533..424O,2001ApJ...562..887T,2005NuPhA.752..550Q,2011PrPNP..66..346T} even though most recent simulations do not yield favorable conditions for the $r$-process \cite{arcones2007,2010A&A...517A..80F,2010PhRvL.104y1101H}. Other candidate sites include compact object mergers \cite{1977ApJ...213..225L,1999ApJ...525L.121F,2005NuPhA.758..587G,2008ApJ...679L.117S,2010MNRAS.406.2650M,2011ApJ...738L..32G,2011arXiv1106.6142W,Caballero2011arXiv1105.6371C}, gamma-ray burst outflows \cite{2004ApJ...603..611S,2005NuPhA.758..189M,2006ApJ...643.1057S}, neutrino induced nucleosynthesis in He shells \cite{PhysRevLett.106.201104,2011PhRvL.106t1104B}, supernova fallback \cite{2006ApJ...646L.131F}, and collapse of O-Ne-Mg cores \cite{1998ApJ...493L.101W,2003ApJ...593..968W,2007ApJ...667L.159N}. 

Experimentally, it is difficult to measure the properties of the short-lived nuclei far from stability that participate in the $r$-process. Recent developments using radioactive beams show promise (e.g. \cite{PhysRevLett.94.112501,2009AIPC.1098..153J}), but current experimental data on neutron-rich isotopes is limited. Thus $r$-process studies must rely not only on model calculations of the environment, but also on theoretical mass models, e.g. \cite{Moller1995,1996PhLB..387..455P,2009PhRvL.102o2503G}.

Despite these difficulties, much has been learned about the $r$-process over the past 50 years. The most prominent features in the $r$-process abundance distribution above atomic mass number of $A=100$ are two distinct peaks occurring at $A=130$ and $A=195$. It was hypothesized very early that the formation of these peaks should be associated with the long $\beta$-decay rates of closed neutron shells \cite{1957RvMP...29..547B}. Since this seminal paper much effort has been put in to researching the conditions for a sufficient initial neutron-to-seed ratio, a key requirement in order to produce a \textquoteleft main\textquoteright\ $r$-process out to the third peak ($A=195$). For reviews see \cite{Arnould200797,2003PrPNP..50..153Q,Cowan:1991zz,Qian2007237}.

\begin{figure*}[htp]
      \includegraphics[width=85mm,height=61.8mm]{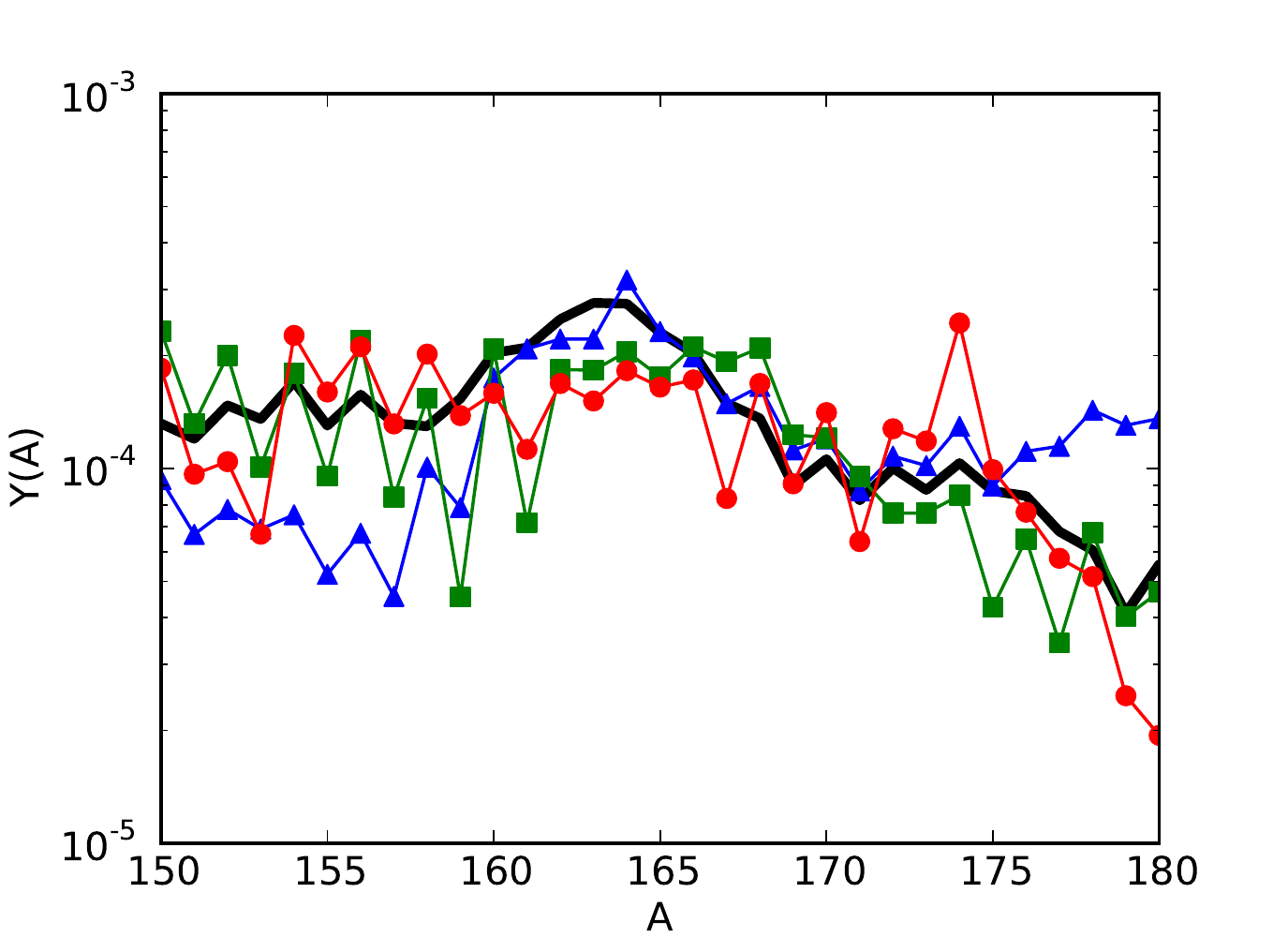}
      \caption{\label{fig:rep} The resultant rare earth peaks from simulations with different mass models. FRDM \cite{Moller1995} (triangles), ETFSI \cite{1996PhLB..387..455P} (squares) and HFB-17 \cite{2009PhRvL.102o2503G} (circles) are shown along with the solar $r$-process abundance pattern, black line, \sdata \ versus atomic mass (data from \cite{1989RPPh...52..945K}). The same colors and geometric markers for each mass model will be used in the remaining figures.}
\end{figure*}

After the two main peaks, the second most prominent feature above A=100 is the smaller peak near $A\sim160$ known as the rare earth peak. While less abundant than the other peaks, the rare earth peak can in principle be used as a powerful tool and offers an alternative way to probe the $r$-process. This is due to the following properties:
(1) Observational data from metal-poor stars show very consistent trends among the rare earth and heavier elements. This suggests that these elements were created in the same type of synthesis event \cite{2008ARA&A..46..241S}. Thus, the rare earth peak provides a natural diagnostic of $r$-process models.
(2) The rare earth peak forms away from closed neutron or proton shells and therefore by a different mechanism than the main peaks. This means it is a different and unique probe of late-time $r$-process conditions.
(3) The rare earth peak is extremely sensitive not only to late-time thermodynamic behavior, but also to nuclear physics input \cite{Surman:1997,Mumpower2010ncr,2011PhRvC..83d5809A}. Typical variations in final rare earth abundance patterns from simulations with different mass models are highlighted in Figure \ref{fig:rep}.

To date the rare earth region has received relatively little attention. Fission cycling has been suggested as a mechanism for obtaining the rare earth peak \cite{1957PASP...69..201C,1971Natur.231..103S}, but it is not favored \cite{marti&zeh1985}. Large uncertainties found in fission probabilities and fragment distributions of current nuclear models further compound difficulties with a successful description of rare earth peak formation by fission cycling \cite{2011ApJ...738L..32G}. Surman et al. \cite{Surman:1997} investigated the formation of the peak in a hot $r$-process environment with temperatures high enough to support \nggn \ equilibrium. The formation of the rare earth peak under these conditions was attributed to the co-action of nuclear deformation and $\beta$-decay as the free neutrons are quickly captured during freeze-out. This was followed by a study of late-time abundances changes among the major peaks \cite{2001PhRvC..64c5801S}. Otsuki et al. \cite{2003NewA....8..767O} investigated a range of r-process models and found similar rare earth elemental abundance patterns, provided the temperature was constant during freeze-out. Most recently, Arcones et al. \cite{2011PhRvC..83d5809A} studied the sensitivity of late-time abundance fluctuations to changes in the nuclear physics inputs. Arcones et al. \cite{2011PhRvC..83d5809A} pointed out that the rare earth peak is sensitive to changes at late-times, e.g. to non-equilibrium effects such as neutron capture even when the abundance of free neutrons can become very low ($\sim10^{-5}$).

This manuscript presents a more complete picture of rare earth peak formation. We explore the sensitivity of the peak formation mechanism to late-time thermodynamic behavior and nuclear physics input. The \textquoteleft funneling\textquoteright\ formation mechanism of \cite{Surman:1997} is reviewed for hot evolutions. We introduce a different \textquoteleft trapping\textquoteright\ mechanism for peak formation in cold evolutions where the temperatures and densities decline relatively quickly and therefore photo-dissociation plays no role in the late-time dynamics after $R=1$. We study the effects of three different mass models and show how large uncertainties in this region stem from nuclear physics. Lastly, we show that the nuclei which contribute to peak formation are approximately 10 to 15 neutrons from stability, and thus represent prime candidates to be measured in future radioactive ion beam facilities (FRIB \footnote{http://www.frib.msu.edu} or FAIR \footnote{http://www.gsi.de/portrait/fair.html}).

\section{r-Process Conditions and Calculations}
Abundance weighted lifetimes are used throughout the text to characterize the late-time dynamics of the $r$-process. These are provided below for the reader's convenience:

\begin{subequations}
\begin{equation}\label{eqn:tau-ng}
\tau_{n\gamma} \equiv \frac{\sum_{Z\geqslant8,A}Y(Z,A)}{\sum_{Z\geqslant8,A}N_{n}\langle \sigma v\rangle_{Z,A} Y(Z,A)}
\end{equation}

\begin{equation}\label{eqn:tau-gn}
\tau_{\gamma n} \equiv \frac{\sum_{Z\geqslant8,A}Y(Z,A)}{\sum_{Z\geqslant8,A}\lambda_{\gamma n}(Z,A)Y(Z,A)}
\end{equation}

\begin{equation}\label{eqn:tau-beta}
\tau_{\beta} \equiv \frac{\sum_{Z\geqslant8,A}Y(Z,A)}{\sum_{Z\geqslant8,A}\lambda_{\beta}(Z,A)Y(Z,A)}
\end{equation}
\end{subequations}

where $N_{n}$ is the neutron number density, $\langle \sigma v\rangle_{Z,A}$ the thermally averaged neutron capture cross section for nuclei $(Z,A)$, $\lambda_{\gamma n}(Z,A)$ the photo-dissociation rate for nuclei $(Z,A)$, $\lambda_{\beta}(Z,A)$ the full $\beta$-decay rate (including $\beta$-delayed neutron emission channels) for nuclei $(Z,A)$ and $Y(Z,A)$ the abundance of nuclei $(Z,A)$.
A reduced sum denoted with a superscript ``REP'' is taken over the rare earth region, $A=150$ to $A=180$, when applicable. The neutron-to-seed ratio or $R$ is defined as:

\begin{equation}\label{eqn:r}
R \equiv \frac{Y_{n}}{\sum_{Z\geqslant8,A}Y(Z,A)}
\end{equation}

where $Y_{n}$ is the abundance of free neutrons.

Since rare earth peak formation is highly dependent on the rate of decrease in the temperature and density, we consider rare earth peak formation under two different thermodynamic evolutions. One scenario is a classical \textquoteleft hot\textquoteright\ $r$-process which operates under high temperatures ($T_{9}\gtrsim1$) at the time in which neutron captures are important for peak formation. A second scenario is a \textquoteleft cold\textquoteright\ $r$-process which operates under low temperatures ($T_{9}\sim0.5$) at the time in which neutron captures are important for peak formation \cite{2007ApJ...666L..77W}.

The classical $r$-process begins with a phase of \nggn \ equilibrium marked by an abundance weighted lifetime ratio of neutron capture to photo-dissociation of \taungtaugn$=1$. During this phase the temperature is still sufficiently high so that neutron captures dominate $\beta$-decays (\taubtaung$\gg1$) and the Saha equation can be used to determine abundances along an isotopic chain \cite{Cowan:1991zz}.

The second phase, known as the freeze-out epoch, is marked by the weakening of the \nggn \ equilibrium  (\taungtaugn$\lesssim1$) and the abundance weighted lifetime ratio of $\beta$-decay versus neutron capture falls to \taubtaung$\approx1$. It is during this phase that the formation of the rare earth peak proceeds with competition between neutron captures, photo-disintegrations and $\beta$-decays.

In the cold $r$-process the first phase \nggn \ equilibrium is dramatically shorter than the first phase of the classical scenario. Freeze-out is now caused by a rapid drop in temperature rather than the consumption of free neutrons (as in the classical case). The bulk of the cold $r$-process operates in the second phase, under low temperatures ($T_{9}\sim0.5$), where photo-disintegrations have frozen out \cite{2007ApJ...666L..77W}. 

Once neutron exhaustion ($R=1$) occurs in the cold $r$-process the free neutrons available to the system must come from the recapture of $\beta$-delayed emitted neutrons. The importance of this effect on the final abundance distribution was noted in \cite{2010ApJ...712.1359F,2011PhRvC..83d5809A}. This recapture effect is crucial to peak formation as can be seen from the fact that malformed abundance distributions result if $\beta$-delayed neutron emission is artificially turned off (see \cite{2011PhRvC..83d5809A}).

Our calculations consists of a nuclear reaction network containing $r$-process relevant nuclides as described in \cite{Surman:1997,2001PhRvC..64c5801S}. Previous versions of this network code have been used in the studies of Beun \cite{Beun:2008gn} and Surman \cite{Surman:2008ef}. The primary reaction channels for nuclides in this section of the reaction network are beta-decay, neutron capture, and photo-dissociation. Our fully implicit $r$-process reaction network handles consistently neutron capture rates at low temperatures and calculations with low abundances of free neutrons, both important for simulations with cold evolutions. For the initial abundances we use self-consistent output from an intermediate reaction network \cite{Hix:1999yd} with PARDISO solver \cite{Schenk2004}.

Our $r$-process calculations start at $T_{9}=2$ with densities $\rho\approx9\cdot10^{8}$ g/cm$^{3}$ for hot evolutions with $Y_{e}=.30$ and $\rho\approx5\cdot10^{8}$ g/cm$^{3}$ for cold evolutions with $Y_{e}=.40$. At this time the neutron-to-seed ratios are $R\approx45$ and $R\approx35$ respectively. We study the late-time hot and cold $r$-process evolutions in the context of a monotonically decreasing temperature with density parameterized as:
\begin{equation}\label{eqn:Rho-late}
\rho(t)\propto t^{-n}
\end{equation}
where $n$ controls the type of late-time $r$-process evolution (the time when rare earth peak formation occurs). For hot $r$-process evolutions we set $n=2$ and for cold $r$-process evolutions we set $n=6$. A decaying density of $n=2$ is characteristic of wind models \cite{2002PhRvL..89w1101M,panov&janka2009} at late times while $n=6$ represents a faster decline.

We use three different mass models in our nucleosynthesis calculations: Finite Range Droplet Model (FRDM) \cite{Moller1995}, Extended Thomas-Fermi with Strutinsky Integral and Quenching (ETFSI-Q) \cite{1996PhLB..387..455P} and version 17 of the Hartree Fock Bogoliubov masses (HFB-17) \cite{2009PhRvL.102o2503G}. The FRDM and ETFSI-Q neutron capture rates are from \cite{2000ADNDT..75....1R} and were computed with the statistical model code NON-SMOKER \cite{1998sese.conf..519R}. The HFB-17 neutron capture rates are from the publicly available Brusslib online-database \footnote{http://www.astro.ulb.ac.be/} and were computed with the statistical model code TALYS \cite{2008A&A...487..767G}, which is also publicly available. The HFB mass model is under constant development and is therefore updated with the latest experimental data and theoretical techniques \footnote{http://www.astro.ulb.ac.be/}. The $\beta$-decay rates used in our $r$-process network come from \cite{2003PhRvC..67e5802M}.

\section{Peak Formation in Hot Environments}
The mechanism for rare earth peak formation in hot environments was first described in \cite{Surman:1997}. We review the basic physical arguments in this section.

Under hot conditions the $r$-process path (time ordered set of most abundant isotopes) traverses the NZ-plane between the line of stability and the neutron drip line. The path is initially constrained by \nggn \ equilibrium and is thus found to lie on a line of constant separation energy via the Saha equation. As the free neutrons are consumed, the path moves back toward stability and \nggn \ equilibrium begins to break down. During this freeze-out from equilibrium, rare earth peak formation can potentially occur.

The necessary and sufficient conditions for peak formation are as follows: (1) a deformation maximum or other nuclear structure effect must produce a kink in the lines of constant neutron separation energy around $A\sim160$, and (2) the $r$-process path must traverse this kink region during freeze-out, before $\beta$-decay takes over in the region. The latter allows for the interplay of neutron capture, photo-dissociation and $\beta$-decay as the $r$-process path crosses the region which contains the separation energy kink.

During peak formation, the $r$-process path moves toward stability at a rate approximately equal to the average $\beta$-decay rate along the path. The separation energy kink causes a corresponding kink in the $r$-process path as material moves through this important region. This provides a mismatch between the $\beta$-decay rates of material below and above the kink. Due to the kink in the path, nuclei below the peak ($A=150$ to $A=158$) are farther from stability and so $\beta$-decay faster than the average nuclei along the path. Since the nuclei below the peak decay faster than the path moves, these nuclei then proceed to capture neutrons in an attempt to return the $r$-process path back to equilibrium. Conversely, due to the kink in the path, nuclei above the peak ($A=168$ to $A=180$) are closer to stability and so $\beta$-decay slower than average along the path. The path therefore moves before these nuclei have a chance to decay and so they photo-dissociate to shift the $r$-process path back to equilibrium. In the peak region ($A=159$ to $A=167$) some nuclei are still in \nggn \ equilibrium which limits the amount of material flowing out of the peak region in either direction. The net result causes material to funnel into the peak region, creating the local maximum.

\begin{figure*}
   \begin{center}
      \includegraphics[width=180mm,height=130mm]{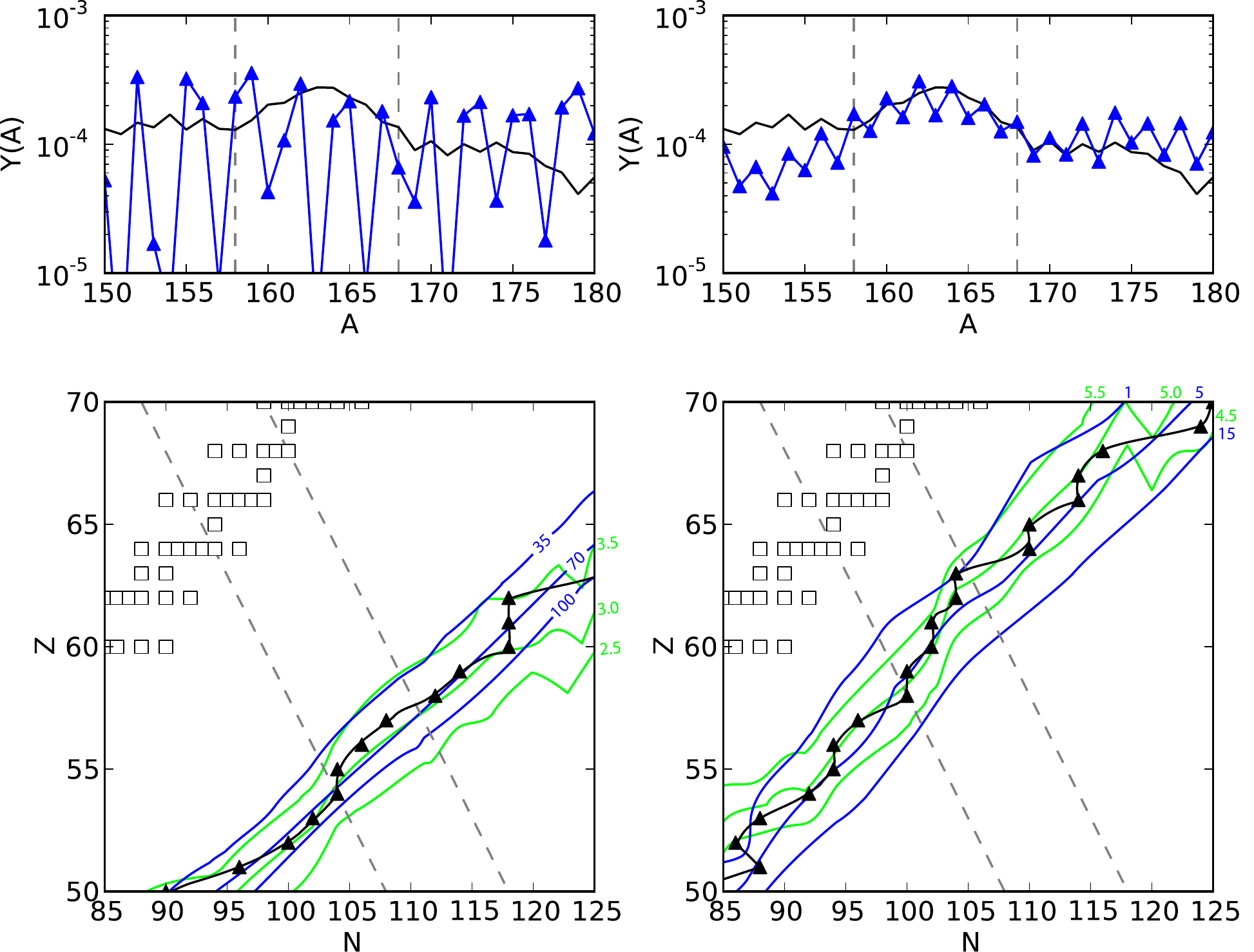}
      \caption{\label{fig:hpf} Shows how the rare earth peak forms under hot environments when the $r$-process path encounters the kink in the separation energies. Each left and right panel shows a snapshot of an abundance pattern along with separation energies (MeV), $\beta$-decay rates ($s^{-1}$), and $r$-process path from a simulation with the FRDM mass model. The left panel is a snapshot at neutron exhaustion, $R=1$. The right panel is a snapshot of the $r$-process path beginning to move closer to stability while the peak forms. Connected black triangles represent the $r$-process path. Blue lines represent constant total $\beta$-decay rate for even-N nuclei. Green lines represent constant separation energy for even-N nuclei. Faint diagonal dotted lines delineate the borders of the peak region ($A=159$ to $A=167$). Stable isotopes are shown by unfilled squares.}
   \end{center}
\end{figure*}

The essence of this effect is shown in Figure \ref{fig:hpf}. At neutron exhaustion, $R=1$ (left panel), the $r$-process is just beginning to break from \nggn \ equilibrium. Here the path lies along a line of constant separation energy ($\sim3.0$ MeV) and the abundances show an odd-even effect due to the population of primarily even-N nuclei in equilibrium. No peak exists at this time.

Later in the simulation (right panel), peak formation occurs as the path encounters the region with the separation energy kink. The separation energy kink causes the kink in the $r$-process path. Nuclei along the path in the peak region have $\beta$-decay rates which range from 1 $s^{-1}$ (above the kink) to 10 $s^{-1}$ (below the kink). The resultant photo-dissociation above the kink and $\beta$-decay followed by neutron capture below the kink causes material to funnel into the peak region.

\section{Peak Formation in Cold Environments}\label{repf:cold}
In the previous section we analyzed rare earth peak formation in hot evolutions and found that photo-dissociation was crucial in peak formation. However, we also find well formed solar-like rare earth peaks in simulations of cold environments where photo-dissociation plays no role in the dynamics after $R=1$.

After $R=1$, the cold $r$-process path is controlled on average by the competition between neutron captures and $\beta$-decays \taubtaung$\approx1$. Locally, over the rare earth region, the exact position of the path is more complicated due to the variation among individual rates.

As the material decays back to stability peak formation will ensue if the path encounters a peak region where neutron capture rates are slow relative to the above and below regions. The essence of the effect is that slow neutron capture rates in the peak region cause a bow (inwards towards stability) in the lines of constant neutron capture rates relative to the lines of constant $\beta$-decay rates thus causing material to become trapped in the peak region.

\begin{figure*}
   \begin{center}
      \includegraphics[width=180mm,height=130mm]{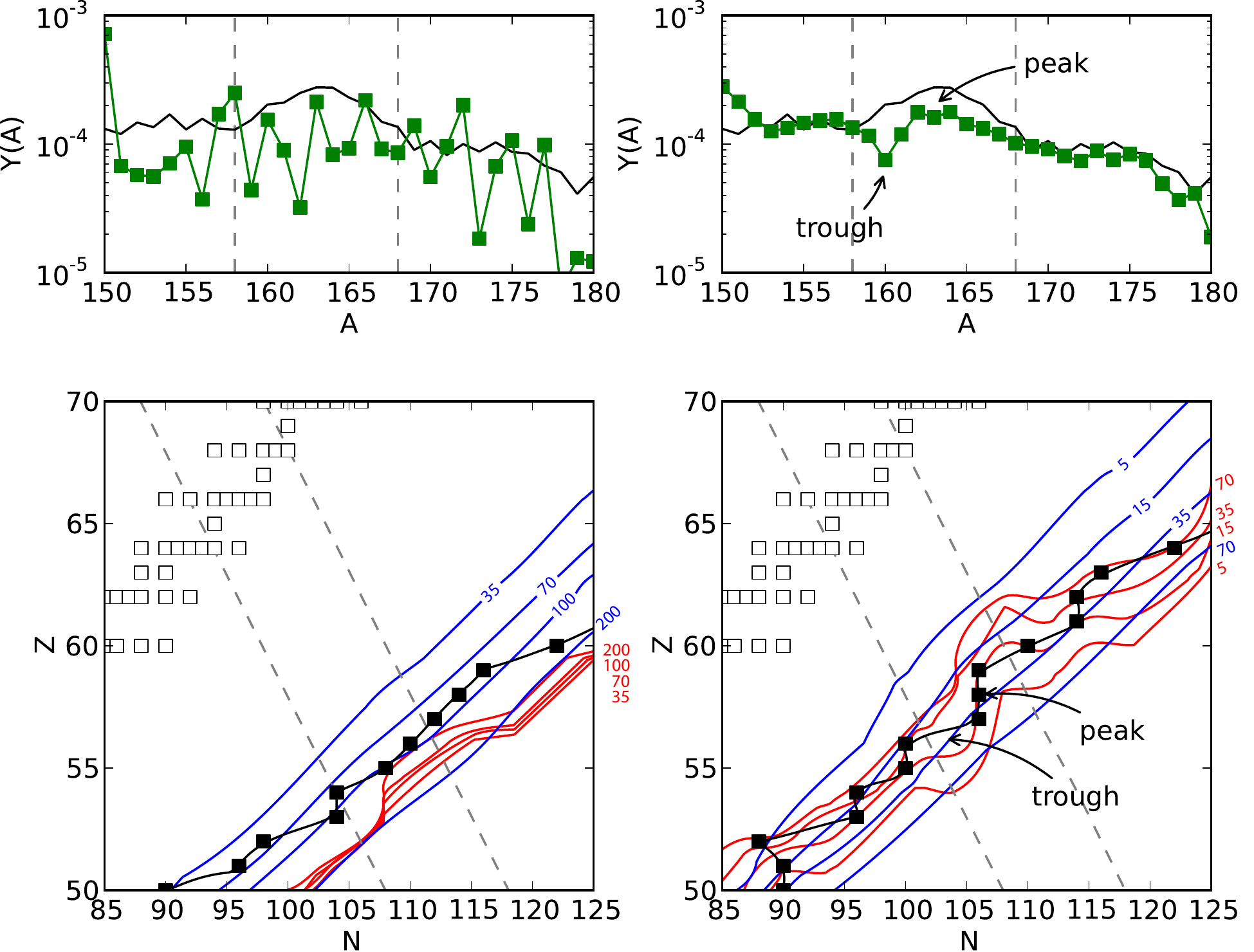}
      \caption{\label{fig:cpf} Shows how the rare earth peak forms under cold environments when the $r$-process path encounters slower neutron capture rates in the peak region. Each left and right panel shows a snapshot of an abundance pattern along with neutron capture rates ($s^{-1}$), $\beta$-decay rates ($s^{-1}$), and $r$-process path from a simulation with the ETFSI mass model. The left panel is a snapshot just after neutron exhaustion, $R=1$. The right panel is a snapshot showing the abundance pattern and rates as the $r$-process path begins to move closer to stability and the peak begins to form. Connected black squares represent the $r$-process path and red lines represent constant neutron capture rates for even-N nuclei. All other markers are the same as Figure \ref{fig:hpf}}
   \end{center}
\end{figure*}

The cold formation mechanism is shown in Figure \ref{fig:cpf}. The left panel shows a snapshot of the abundance pattern and rates at neutron exhaustion, $R=1$. At this point in time the $r$-process path is still influenced by residual photo-dissociation flows. This is reflected in an odd-even effect in the abundances and flat $r$-process path (similar to hot evolutions). However, the photo-dissociation rates are decreasing so rapidly they play no further role in the dynamics after this point. Shortly, the neutron capture rates will become comparable to the $\beta$-decay rates and large odd-even behavior of the abundances will be washed-out \cite{1999ApJ...516..381F}. In fact, this has already begun to happen as can be seen with the slight bowing of the neutron capture rate lines in the peak region ($A=159$ to $A=167$).

At a slightly later time in the simulation (right panel of Figure \ref{fig:cpf}) the system has moved closer to stability and the $r$-process path now encounters the slower capture rates in the peak region. Below the peak ($A=150$ to $A=158$) neutron captures occur much faster than $\beta$-decay rates along the $r$-process path, so the net result is material shifting towards the peak region. In the peak region the path encounters the slow capture rates (note the bowing of the neutron capture rate lines) so that any material being shifted into the peak region becomes hung up. Above the peak ($A=168$ to $A=180$) the flow of material is again dominated by the relatively faster neutron capture rates. The net result is trapping of material into the peak region.

Another interesting feature in the right panel of Figure \ref{fig:cpf} is the trough to the left of the peak. A trough can occur if a gap in the $r$-process path proceeds for long periods of time as matter decays back to stability. Along a gap in the $r$-process path the neutron capture rates are relatively fast resulting in movement of material to more neutron-rich isotopes and a depletion of material in the gap region.

In our figures, the lines of constant neutron capture rates have been averaged over even-N nuclei. Even-N neutron capture rates are more important to rare earth peak formation because at a given temperature, odd-N nuclei have faster neutron capture rates which causes material to pass through the odd-N nuclei quickly. Thus material builds up (or stays) in even-N nuclei which sets the $r$-process path. The importance of individual neutron capture rates in the rare earth peak was highlighted in \cite{Mumpower2010ncr}.

\section{Influence of Mass Model on Rare Earth Peak Formation}
From the previous two sections it is clear that the details of the late-time thermodynamic evolution are critical in setting the relevant nuclear physics and thus determine the mechanism for peak formation.

Despite the differences in peak formation mechanisms, we find that the final abundances among simulations with the same mass model yet differing late-time thermodynamic behavior can be remarkably similar. This is in contrast to the differences found in the final abundance pattern when comparing between mass models with similar thermodynamic conditions. In this section we focus on the influence of different separation energies and neutron capture rates on rare earth peak formation.

A successful peak formation is imprinted on the final abundances in a cold evolution when the $r$-process path encounters structure in the neutron capture rates and this structure lasts until the point at which $\beta$-decays take over neutron captures in the region (\reptaungafewb).

A successful peak formation occurs in a hot evolution when the $r$-process path encounters a local deformation maximum leading to a well-defined kink structure in the separation energies in the rare earth region.

For a given mass model, the structure of neutron capture rates and the structure of the separation energies may not align in the NZ-plane. This in turn can affect the timing and location of peak formation and hence the nuclei which are relevant.

Odd-even effects in the abundances can accumulate or persist through the decay back to stability resulting in visible features in the final abundances. Smoothing of the abundances typically occurs in between neutron capture freeze-out (\reptaungeqb) and the time in which $\beta$-decays fully take over neutron captures in the region (\reptaungafewb).

We now discuss three different mass models in this context. Since separation energies vary among mass models we instead (for consistency) use $\langle\delta N\rangle$, the abundance weighted average neutrons from stability, to measure the $r$-process path's progression.

Compared to the other mass models studied here, we find that simulations which use the FRDM mass model best match the solar data in the rare earth peak region in both hot and cold evolutions. In fact we find (in agreement with previous studies \cite{Surman:1997,2011PhRvC..83d5809A}) that the FRDM mass model is the only model to show a well-formed rare earth peak consistently in the final abundance pattern.

\begin{figure*}
   \begin{center}
      \includegraphics[width=180mm,height=140mm]{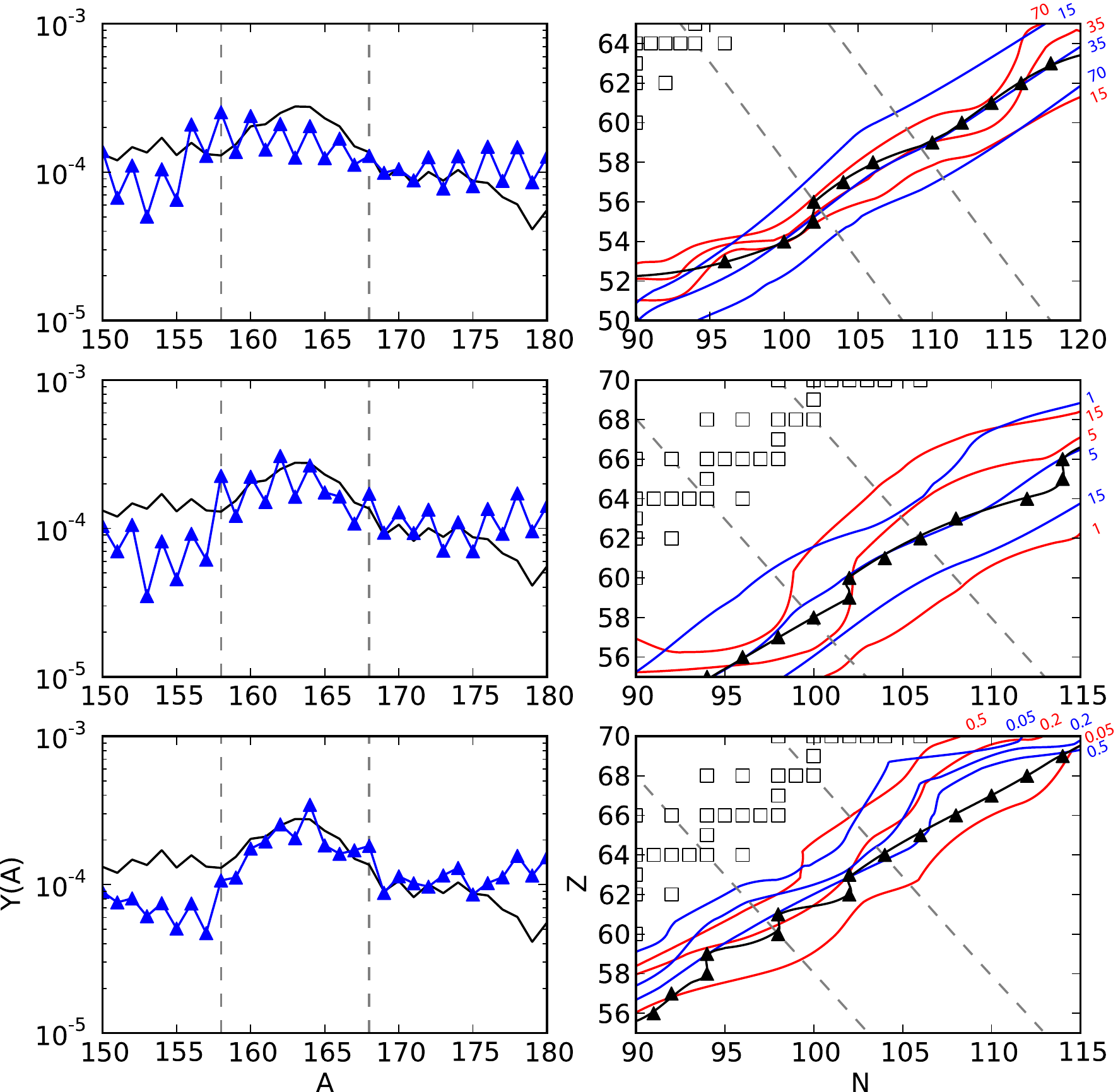}
      \caption{\label{fig:form-cold-frdm} A successful rare earth peak formation under a cold evolution with FRDM mass model occurs when the structure in the neutron capture rates lasts until the point at which $\beta$-decays take over neutron captures in the region (\reptaungafewb). Neutron capture rates (online red, print dotted), $\beta$-decay rates (online blue, print solid), $r$-process path (filled triangles in the right column) and abundance snapshots (left column) are shown at 20 (top panel), 15 (middle panel) and 10 (bottom panel) neutrons away from stability as the peak evolves during late-times in this cold FRDM simulation. The structure in the neutron capture rates is not yet present at 20 neutrons away from stability (top panel). However, the structure in the neutron capture rates becomes evident at 15 neutrons away from stability (middle panel), continuing until the abundance pattern has frozen out completely. The bottom panel abundances are also highlighted in Figure \ref{fig:rep}.}
   \end{center}
\end{figure*}

\begin{figure*}
   \begin{center}
      \includegraphics[width=180mm,height=140mm]{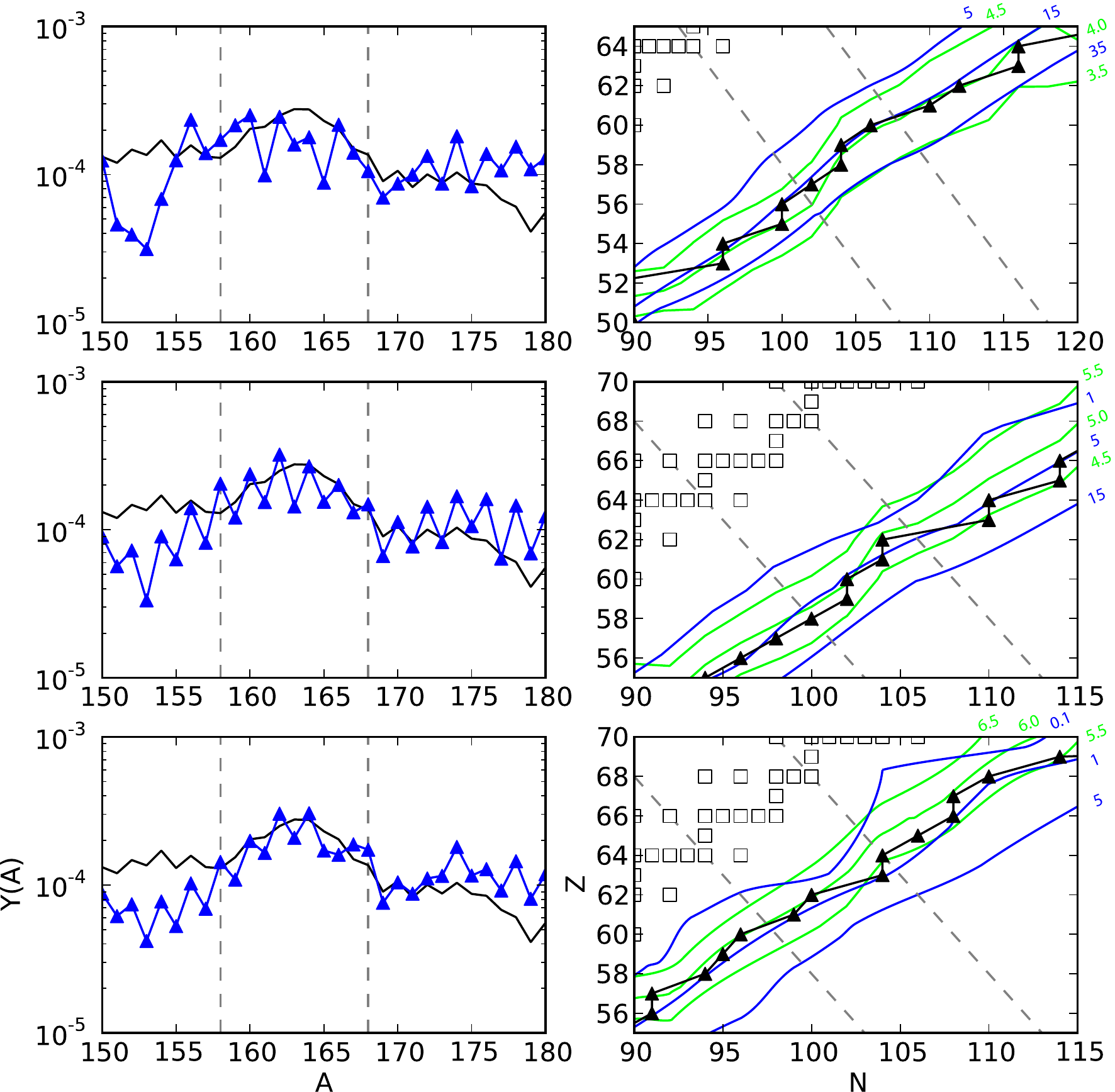}
      \caption{\label{fig:form-hot-frdm} A successful rare earth peak formation under a hot evolution with FRDM mass model. The $r$-process path encounters the well-defined separation energy kink in between 20 (top panel) and 15 (middle panel) neutrons away from stability in the peak region. The separation energy kink does not last (bottom panel) but by this time $\beta$-decays have taken over and the abundances have frozen-out. Separation energies (online green, print dotted), $\beta$-decay rates (online blue, print solid), $r$-process path (filled triangles in the right column) and abundance snapshots (left column) are shown at 20 (top panel), 15 (middle panel) and 10 (bottom panel) neutrons away from stability similar to the previous figure.}
   \end{center}
\end{figure*}

Simulations with the FRDM mass model do not consistently form rare earth peaks far from stability ($\langle\delta N\rangle>20$). Instead, peak formation ensues when the path is much closer; on average in between 15 and 20 neutrons away from stability. We can see the evolution of the peak region for a cold FRDM evolution in Figure \ref{fig:form-cold-frdm}. At $\langle\delta N\rangle\sim20$ (top panel) the structure in the capture rates has yet to manifest itself resulting in relatively flat abundances. As the path moves back to stability, $\langle\delta N\rangle\sim15$ (middle panel), it encounters nuclei in the peak region with relatively slower neutron capture rates than the surrounding regions (note the bending in the red lines). These conditions persist all the way back to stability resulting in a well-formed rare earth peak. A similar scenario occurs in hot evolutions; see Figure \ref{fig:form-hot-frdm}.

FRDM shows a slight overlap between neutron capture structure and separation energy structure. The structure in the separation energies occurring between $\langle\delta N\rangle\sim12\text{ to }20$ and the structure in the capture rates occurring between $\langle\delta N\rangle\sim10\text{ to }15$. This delays peak formation in cold scenarios until around 15 neutrons from stability, while hot evolutions typically begin peak formation approximately 20 neutrons from stability.

\begin{figure*}
   \begin{center}
      \includegraphics[width=180mm,height=140mm]{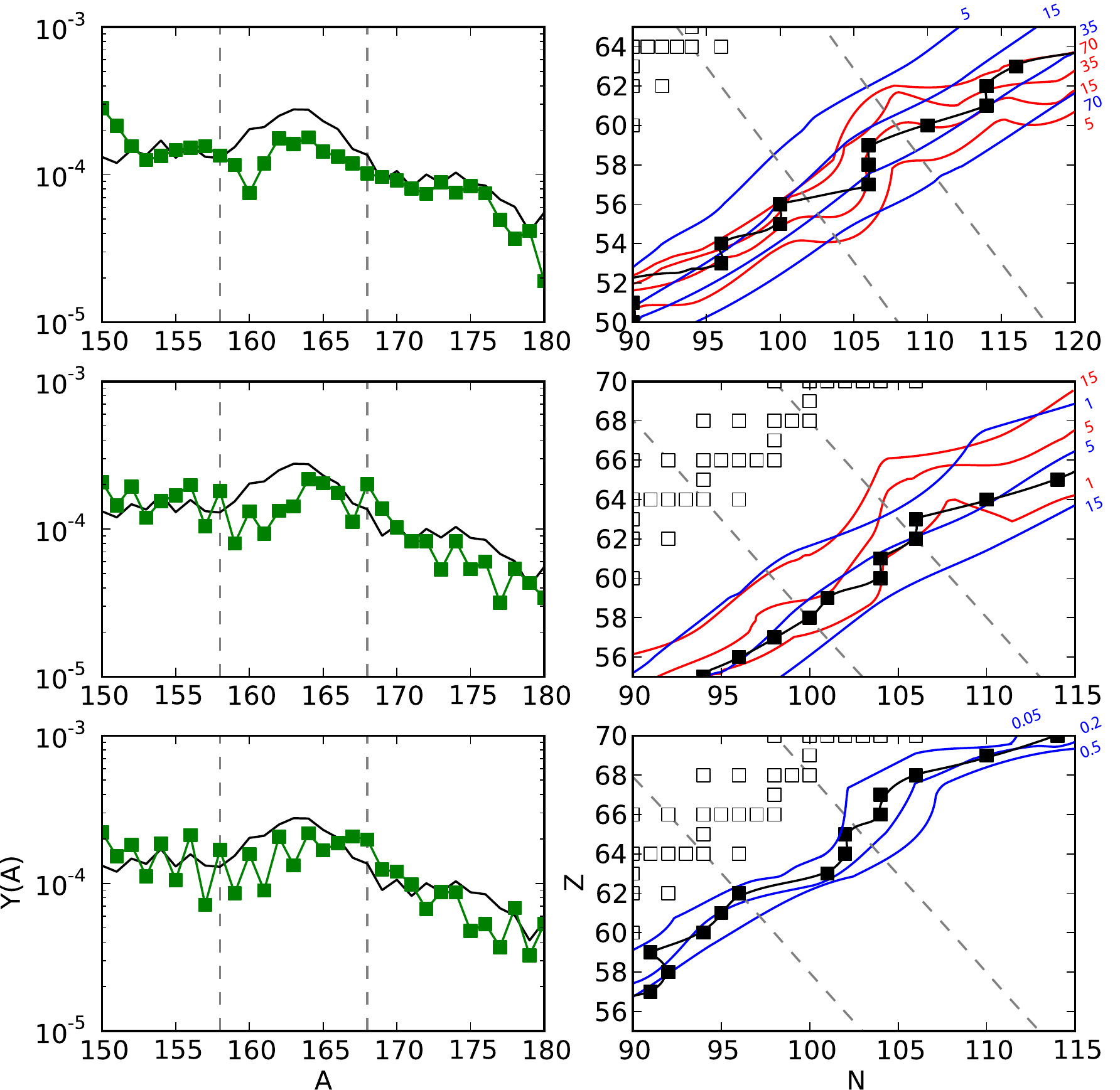}
      \caption{\label{fig:form-cold-etfsi} The rare earth peak forms far from stability in this cold ETFSI simulation (top panel). However, it is washed out by the slow reduction in the structure of neutron capture rates of nuclei closer to stability (middle panel). At 10 neutrons away from stability (bottom panel) the neutron capture rates are slower than $\beta$-decay rates, so that only small changes to the abundance pattern occur after this point. Hence the disappearance of the lines of constant neutron capture rates in the bottom panel. Neutron capture rates (online red, print dotted), $\beta$-decay rates (online blue, print solid), $r$-process path (filled squares in the right column) and abundance snapshots (left column) are shown at 20 (top panel), 15 (middle panel) and 10 (bottom panel) neutrons away from stability as in previous figures.}
   \end{center}
\end{figure*}

Simulations with the ETFSI mass model consistently form a solar-like rare earth peak far from the stable nuclei ($\langle\delta N\rangle\gtrsim20$). This is most apparent in colder simulations (see right panel of Figure \ref{fig:cpf}). However, this is not the end of the story as the material must decay back to stability. Figure \ref{fig:form-cold-etfsi} highlights this transition at an abundance weighted average of $\langle\delta N\rangle\sim20$ (top panel), 15 (middle panel) and 10 (bottom panel) neutrons from stability. As the decay back to stability proceeds the $r$-process path encounters nuclei whose neutron capture rates become homogeneous around the peak region. This slowly dissolves the structure, flattening the lines of constant neutron capture rates (compare top and middle panels). By the time the path is on average 15 neutrons away from stability (middle panel) the cold trapping mechanism can not continue because neutron capture rates in the peak region are no longer slower than the surrounding regions. These conditions persist back to stability resulting in a final abundance pattern with a more modest rare earth peak. Note that a small odd-even effect reappears since in this model $\beta$-delayed neutron emission is still relevant as neutron capture freezes out in the rare earth region (\reptaungeqb).

\begin{figure*}
   \begin{center}
      \includegraphics[width=180mm,height=140mm]{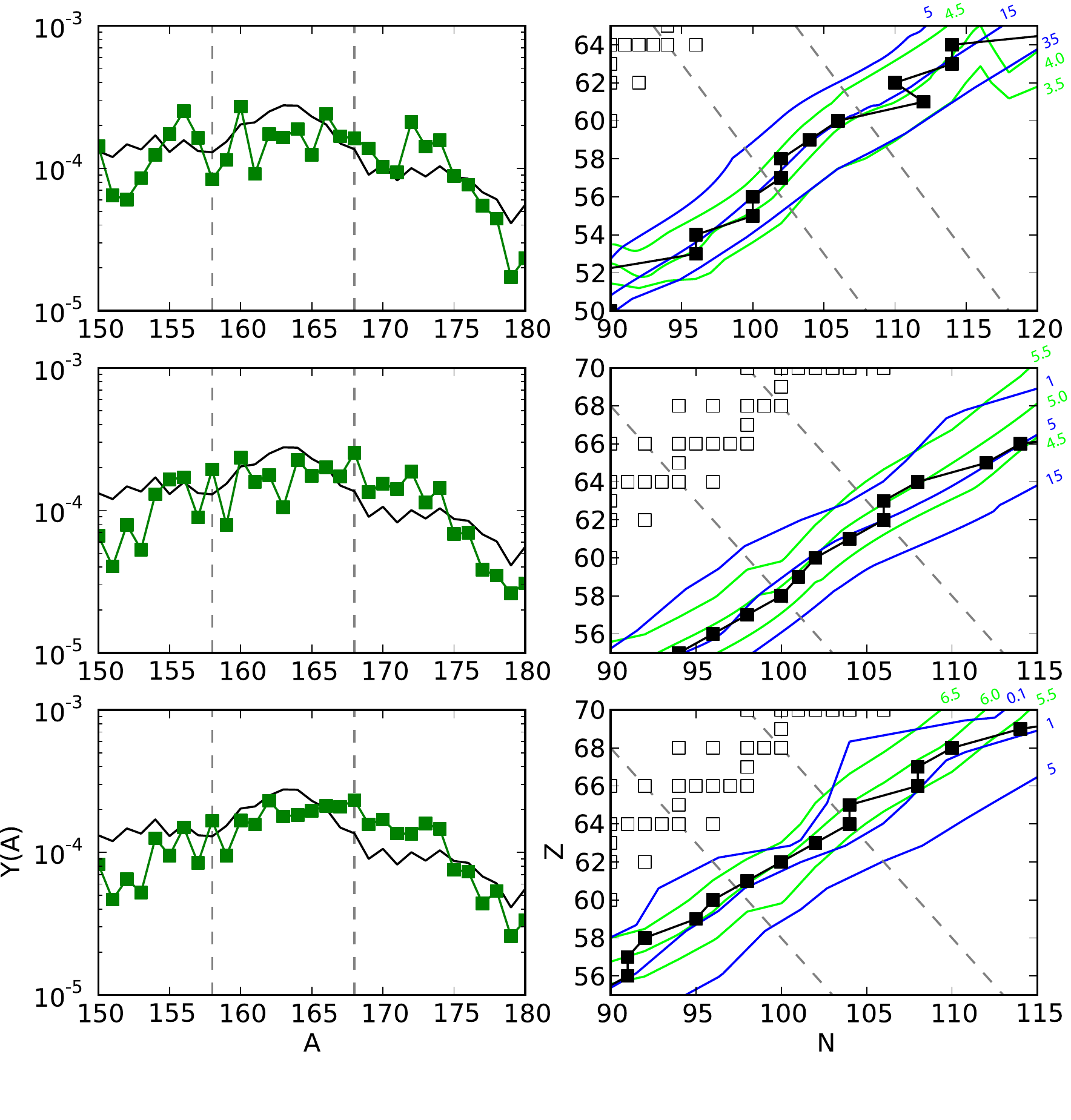}
      \caption{\label{fig:form-hot-etfsi} The rare earth peak begins to form far from stability in this hot ETFSI simulation (top panel) due to the separation energy kink structure beyond 20 neutrons from stability (not shown). However, it is washed out by the slow reduction in the kink structure of the separation energies of nuclei closer to stability (middle panel). Separation energies (online green, print dotted), $\beta$-decay rates (online blue, print solid), $r$-process path (filled squares in the right column) and abundance snapshots (left column) are shown at 20 (top panel), 15 (middle panel) and 10 (bottom panel) neutrons away from stability as in previous figures.}
   \end{center}
\end{figure*}

Solar-like rare earth peaks form far from the stable nuclei in ETFSI models under hot evolutions as well. Far from stability, the structure (kink) in the separation energies results in the hot peak formation mechanism. Like the FRDM case, the separation energy kink in ETFSI disappears as one moves closer to stability. However, the kink disappears while neutron captures are still dominant (\reptaungltb) far from stability ($\langle\delta N\rangle\geqslant20$) resulting in a flattened final abundance distribution; see Figure \ref{fig:form-hot-etfsi}.

In this mass model the structure in the separation energies occurs farther from stability ($\langle\delta N\rangle\geqslant20$) than the structure seen in the neutron capture rates ($\langle\delta N\rangle\sim20$) influencing peak formation in a similar fashion to the FRDM case. The gross separation energy structure occurs very early on ``before'' the top panel of Figure \ref{fig:form-hot-etfsi} and has already dissolved by $\langle\delta N\rangle\sim20$.

Version 17 of the HFB mass model is optimized to over 2000 measured masses from \cite{2003NuPhA.729....3A} corresponding to a root mean square error of $\lesssim0.6$MeV. This data set features detailed structure in the separation energies but little overall structure in the neutron capture rates for the nuclei relevant to rare earth peak formation. These features are reflected in our $r$-process abundances.

\begin{figure*}
   \begin{center}
      \includegraphics[width=180mm,height=140mm]{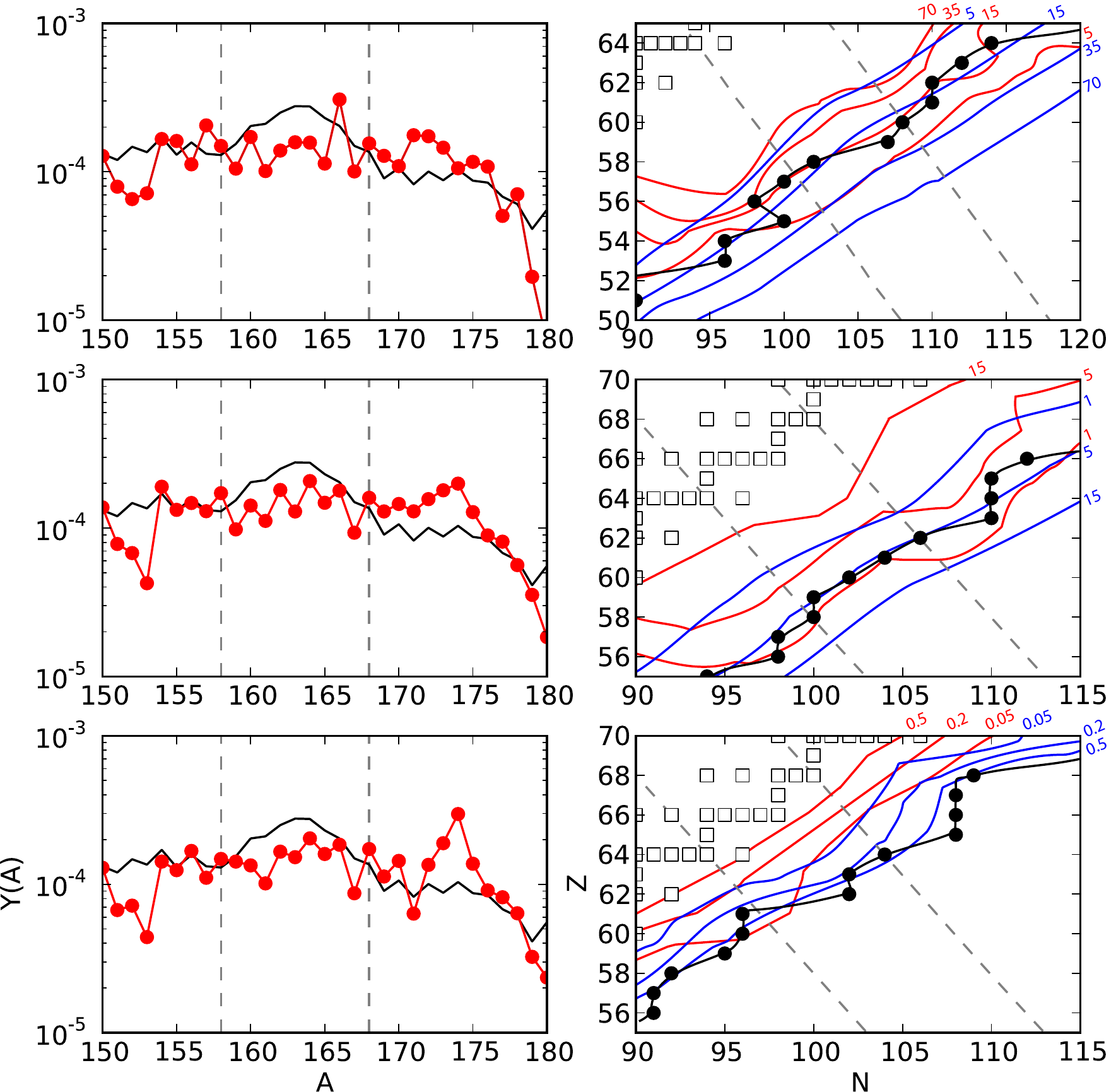}
      \caption{\label{fig:form-cold-hfb17} The final rare earth peak (bottom panel) is relatively flat (as compared to the solar rare earth peak) in this cold HFB-17 simulation. This occurs when the $r$-process path encounters a lack of neutron capture rate structure throughout the NZ-plane in the peak region. Neutron capture rates (online red, print dotted), $\beta$-decay rates (online blue, print solid), $r$-process path (filled circles in the right column) and abundance snapshots (left column) are shown at 20 (top panel), 15 (middle panel) and 10 (bottom panel) neutrons away from stability as in previous figures.}
   \end{center}
\end{figure*}

Figure \ref{fig:form-cold-hfb17} shows the decay back to stability of a cold $r$-process using HFB-17. At every snapshot, highlighting the $r$-process path's decay back to stability, we do not find the structure in the neutron capture rates as is found in the other two mass models. It is this relative homogeneity in the neutron capture rates throughout the rare earth region which prevents the trapping mechanism from occurring in cold evolutions.

\begin{figure*}
   \begin{center}
      \includegraphics[width=180mm,height=140mm]{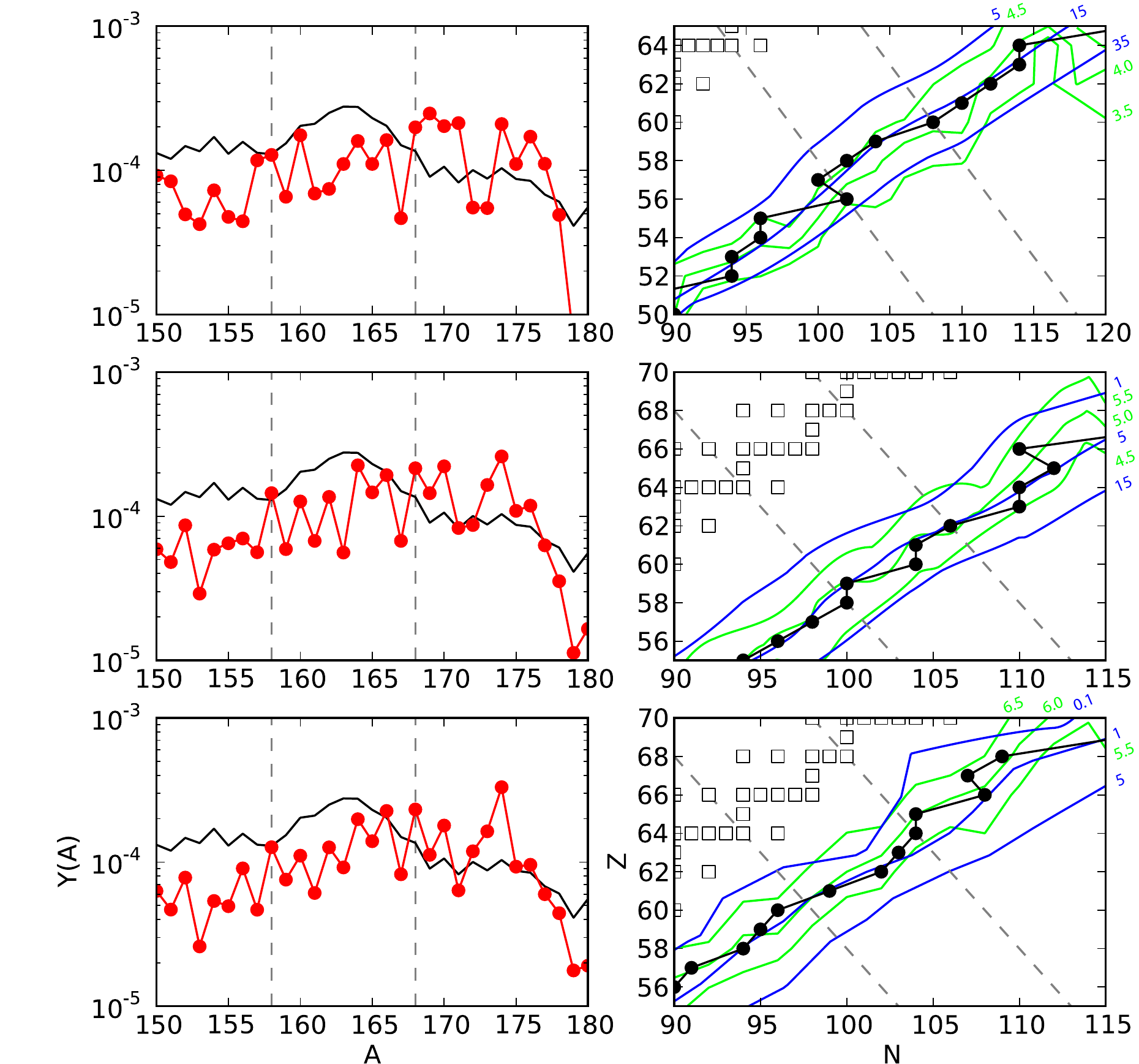}
      \caption{\label{fig:form-hot-hfb17} The final rare earth peak (bottom panel) is relatively flat (as compared to the solar rare earth peak) in this hot HFB-17 simulation. While there is detailed structure in the separation energies, there is little gross structure on the scale of the rare earth peak. The slight abundance bump off-center of the actual peak region is due to the complex structure in the separation energies found off-center from the peak region. Separation energies (online green, print dotted), $\beta$-decay rates (online blue, print solid), $r$-process path (filled circles in the right column) and abundance snapshots (left column) are shown at 20 (top panel), 15 (middle panel) and 10 (bottom panel) neutrons away from stability as in previous figures.}
   \end{center}
\end{figure*}

In hot $r$-process evolutions the situation is more intricate than for the corresponding cases of the other two mass models. The detailed structure in the separation energies results in a complex separation energy kink structure in the rare earth region. However, due to the lack of gross structure as the separation energy increases (i.e. during the decay back to stability) the funneling mechanism cannot operate. This can be seen in Figure \ref{fig:form-hot-hfb17} and illustrates the subtleties involved in forming the rare earth peak.

\begin{figure*}
   \begin{center}
      \includegraphics[width=100mm,height=90mm]{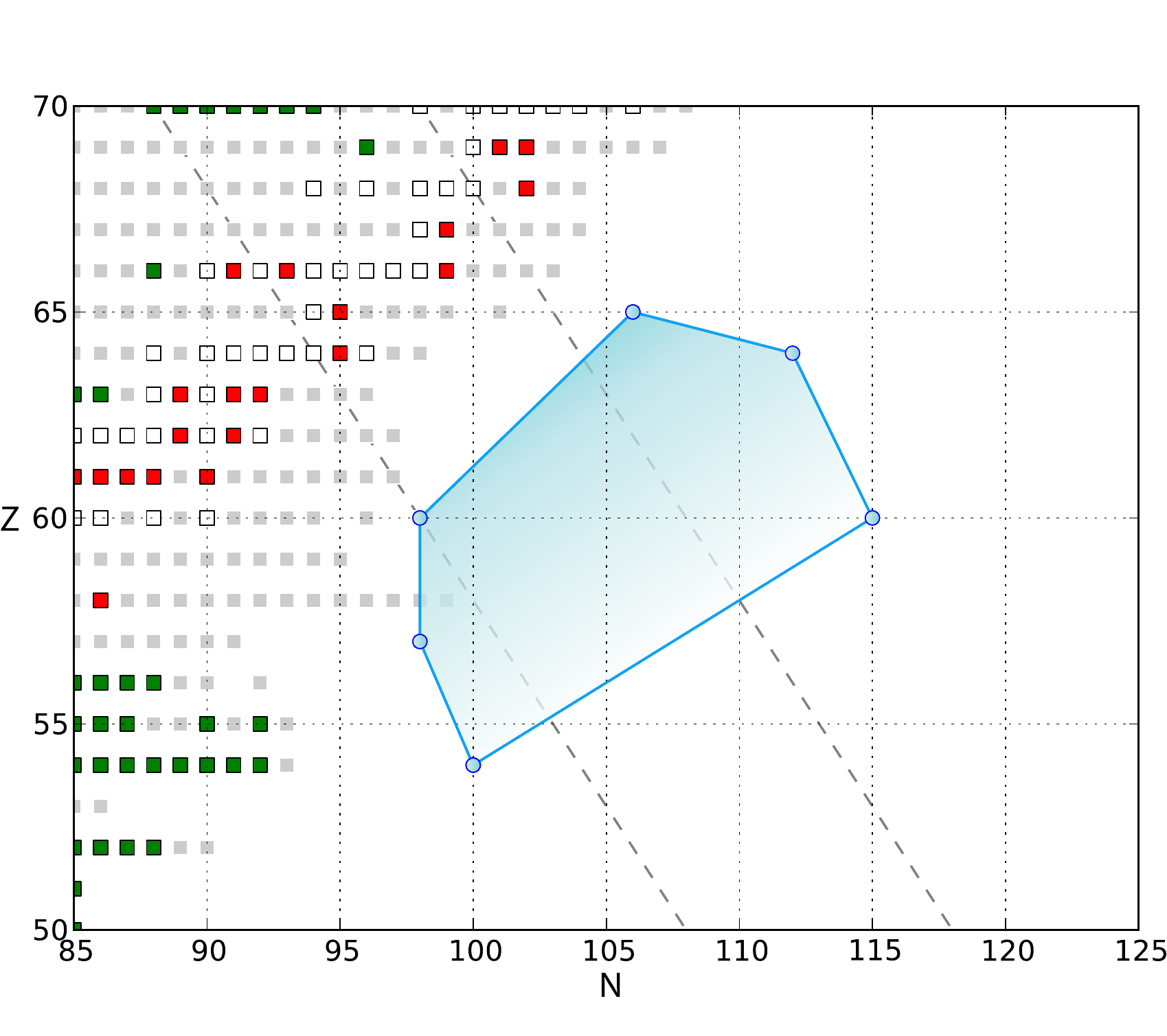}
      \caption{\label{fig:rep-in} Highlights the nuclei which are important to rare earth peak formation. The most influential nuclei are closer to stability in darker shading as they set or potentially dissolve the peak structure. Also shown is the current extent of experimental data for the rare earth elements. Isotopes with measured masses in the AME2003 mass table are highlighted (online and print shaded in light gray). Recent ISOLTRAP and JFYLTRAP mass measurements are highlighted (online green, print dark gray) and cross section data from the online CSISRS database are highlighted (online red, print black). Stable isotopes are shown by unfilled squares as in previous figures.}
   \end{center}
\end{figure*}

The discussion in this section showcases the need for nuclear structure measurements far from stability. As we have seen, the nuclei that are important for rare earth peak formation lie in between 10 and 20 neutrons away from stability. Furthermore, it is the nuclei which are the closest to stability, those in between 10 and 15 neutrons from stability, which are most influential to peak formation as they set or potentially dissolve the peak structure all together. In Figure \ref{fig:rep-in} we highlight these influential nuclei together with recent experimental mass measurements (\footnote{http://isoltrap.web.cern.ch} \footnote{https://www.jyu.fi/fysiikka/en/research/accelerator/igisol/trap} green and \cite{2003NuPhA.729....3A} gray) and known neutron capture rates (\footnote{http://www.nndc.bnl.gov/exfor} red).

\section{Summary and Conclusions}
We have studied the formation and evolution of the rare earth peak at late-times during the $r$-process. To take into account uncertainties with nuclear physics in the region our calculations employed three mass models (FRDM, ETFSI, and HFB-17).

Two late-time evolutions were considered: A hot $r$-process with temperatures high enough to support \nggn \ equilibrium and a cold $r$-process with lower temperatures where there are no photo-dissociation flows, only competition between neutron captures and $\beta$-decays after $R=1$. Both of these evolutions are similar at early times so that the changes in abundances are not due to to the physics that sets the neutron-to-seed ratio, but instead due to the changes in the nuclear physics input or changes in the late-time behavior of the evolution. The differences in late-time evolution (hot vs cold) determine which nuclear physics input is important (separation energies vs neutron capture rates respectively) during the final stages of the $r$-process.

In hot evolutions the combination of photo-dissociation, neutron capture and beta-decay results in a mechanism which funnels material into the peak region. A successful peak formation in hot evolutions is imprinted on the abundance pattern when the structure in the separation energies, the \textquoteleft kink\textquoteright, is well defined \textit{and} the $r$-process path crosses the kink region during the \nggn \ freeze-out.

We contrast this with the peak formation mechanism which occurs in cold $r$-process environments. Here the important nuclear physics for peak formation lies in the local structure of the neutron capture rates. When the neutron capture rates are slow in the peak region relative to the surrounding regions (creating the characteristic \textquoteleft bow\textquoteright\ in the lines of constant neutron capture rates) material can become trapped in the peak region, thus forming the peak. A successful peak formation in cold evolutions is imprinted on the abundance pattern when the structure in the neutron capture rates lasts until the point at which $\beta$-decays take over neutron captures in the region (\reptaungafewb).

The rare earth peak is extremely sensitive to the subtleties of nuclear physics input. Neutron capture is particularly important in both hot and cold evolutions. For instance, we find that neutron capture can play two competing roles in peak formation: it can be responsible for creating the peak, but also for potentially dissolving the peak (wash-out). Neutron capture rate structure and separation energy structure in the same mass model may not overlap in the NZ-plane. This in turn can affect the timing and location of peak formation in different thermodynamic conditions.

We have shown that the rare earth peak in principle offers unique insight into the late-time behavior of the $r$-process because it forms away from the closed shells during freeze-out while material decays back to stability. Rare earth peak formation is sensitive to the structure of separation energies and / or neutron capture rates about 10 to 15 neutrons away from the stable rare earth peak. Future measurements at radioactive ion beam facilities should reach this important region and will be critical in placing constraints on nuclear models. This in turn will lead to improved $r$-process predictions; allowing the rare earth peak to evolve into a powerful tool for understanding the $r$-process.

\section{Acknowledgements}
We thank A. Arcones and T. Rauscher for valuable discussions.
We thank North Carolina State University for providing the high performance computational resources necessary for this project.
This work was supported in part by U.S. DOE Grant No. DE-FG02-02ER41216, DE-SC0004786, and DE-FG02-05ER41398.

\bibliographystyle{unsrt}
\bibliography{repf}


\newpage

\end{document}